# HCI Models for Digital Musical Instruments: Methodologies for Rigorous Testing of Digital Musical Instruments

Gareth W. Young[1] and Dave Murphy[2]

University College Cork
Cork, Ireland.
[1] g.young@cs.ucc.ie
[2] d.murphy@cs.ucc.ie

**Abstract.** Here we present an analysis of literature relating to the evaluation methodologies of Digital Musical Instruments (DMIs) derived from the field of Human Computer Interaction (HCI). We then apply choice aspects from these existing evaluation models and apply them to an optimized evaluation for assessing new DMIs.

## 1 Introduction

The evaluation of computer interface devices in HCI is a well-documented and established topic. There are a number of established and validated HCI evaluation techniques, however none can be said to be fully compatible with respect to DMIs. User focused assessment is an integral part of an interface designer's requirement to quantify and evaluate their technology. Recent developments in user studies have shown an interest in the relationships that users develop with technology and the overall user experience. Previous research has neglected to incorporate and amalgamate these vital aspects in their approach to DMI evaluation. As this field is in a constant state of change, we demonstrate how specific aspects of the aforementioned evaluations can be incorporated into existing DMI evaluation strategies and how they can be applied to current DMI designs.

HCI is a highly complex multivariate discipline, which lacks an all-encompassing device evaluation framework, so we pose the question: in this context, is it possible to accurately evaluate a musical device? A number of researchers have endeavoured to answer this question in reference to DMI design and appraisal, sparking discussion about their proposed methodologies of measurement and if indeed the performance of a DMI may be quantifiably measured at all. Further to this, examples of applied case studies are few, and it appears that designers are cautious to take up and apply these models of analysis to their own experimental devices. Here we shall discuss some aspects of current and proposed HCI evaluation methods for DMIs, and their application to prototype devices.

## 2 Background

In HCI, a number of tools have been developed to measure design parameters, and the use of computers in specific contexts. These tools serve to direct interface designers away from generic, single purpose, interface-testing methods. In this vein, we find ourselves as DMI designers in a HCI context. We can observe this in the techniques that are applied in DMI product design, which are informed through design practices and HCI research. Thusly, a strong connection can be draw between the traditions of HCI and DMI evaluation.

Functionality, usability, and user experience are evaluated in HCI studies in order to create a comprehensive representation of a device in use [1, 2]. For example, when playing music on a basic MIDI keyboard, many will agree that, in general, the usability of the interface is poor in comparison to that of performing on a grand piano. However, the experience may still remain believable or natural for the performer. Additionally, different manufacturers incorporate various additional features in their products in order to attract potential customers with differing requirements. For these reasons, we suggest it may be possible to evaluate a DMI in terms of the general area of its technology usage. Specifically, we recommend evaluating DMI devices in terms of functionality, usability, and the user's experience using it, which are an integral part in our proposed evaluation framework.

Problems arise in DMI evaluation when we consider the wide range of variables involved in musical performance. For live performances of computer music there are a multitude of contributing factors to a musician's experience, these include: the consideration of simultaneous timing and rhythmic patterns, a performer's previous training with a specific instrument and their familiarity with other instruments within a collective ensemble. Coupled with this we need to consider the multi-parametric control afforded at different levels, which are dependent upon the mechanical characteristics of the chosen instrument. Proposals have been made in the past to make a quantifiable and comparative analysis of devices over a series of short representative tasks. Additionally, the categorization of input devices to match tasks has also been suggested in order to adhere to specific and measurable objectives that match the operational characteristics of the individual device.

In order to appraise all critical aspects of a DMI, we must assess each evaluation area closely for its applicability to the chosen device. There may also be reason to assess one-off DMIs with unique and augmentable sets of evaluation methods in order to achieve this. Therefore, it is important for us to firstly acknowledge that any investigation of a DMI's design may incorporate its own set of unique methodologies and assumptions, highlighting the necessity to carefully choose approaches that best fit the device for the three evaluation areas outlined earlier. For example, the appraisal of standard Usability Evaluation Methods (UEMs), such as time-on-task and number-of-errors for instance, cannot be used alone to assess a user's experience. Similarly, UEMs used to assess a device's functionality are not sufficient. In order for an accurate appraisal of a device, we must be careful not to reduce our analysis to any rigid or single base form.

## 3 Previous Research Findings

Notable examples of crossover between HCI-DMI evaluation methods can be seen in a number of previous publications. Research focused on the adaption of existing HCI tools and methods have been identified [3]. However, in practice widespread use of these HCI-DMI crossovers are limited to just a few examples [4]. Orio et al. draw together some of the most appropriate evaluation methods that apply to musical devices and discusses these in a musical context. They highlight target acquisition as a potential quantifiable evaluation method, underlining Fitts' Law and Meyer's Law in particular. In a musical context, consideration of *learnability, explorability, feature controllability*, and *timing controllability* were also emphasized as important aspects in the classification of controllers in terms of their usability [3]. The mechanical characteristics of a DMI were also highlighted as having a categorical impact on device comparisons. Matching devices with similar, basic characteristics, or taxonomies is imperative for an organized and fair comparison.

In order to organize DMI classifications, there have been a number of potential guidelines published. With the propagation of new interfaces for musical expression in digital music, it has been noted that the application of hardware interfaces, control surfaces, and gesture-based controllers are of considerable interest to musicians. The classification of custom devices for musical application has also received increased attention. Miranda and Wanderley propose several distinct categories of DMI [5]. Their categorization includes: *instrument-like controllers*, *extended instruments*, *instrument-inspired controllers*, and *alternative controllers*. Upon further examination, two major distinctions can be made between these groups. For *instrument-like controllers*, *extended instruments*, and *instrument-inspired controllers*, the instrument designer is restricted to a musician's musically refined motor control ability or familiarity of an instrument's mode of interaction, which is either practiced or is inherently familiar. In many *alternative controllers*, this familiarity is actively avoided, allowing for the use of non-traditional gesture vocabularies to be explored by a performer. Additionally, as the designer, composer, and performer may be the same person, the design of the instrument may be very individual, which makes it difficult when formally assessing the device's performance as a DMI.

Wanderley and Orio further expand on their earlier findings by introducing contextual events to use when comparing categorized devices. The expansion of categorical comparison was achieved by presenting an expanded list of circumstances specific to interactive computer music [6]. The contexts in which these categories are applied include: *note-level control* or *musical instrument manipulation, score-level control, sound processing control* or *post-production activities, context related to traditional HCI or navigation,* and *interaction in multimedia installations*. Additionally, the authors saw fit to include metaphoric situations, where generating music was not necessarily the primary focus of the interaction, such as: *interaction in the context of dance/music interfaces,* and *control of computer games*. These classifications were intended to assist in analysis and were not to be considered as fixed. The application of a single device in multiple contexts was also considered an important and distinguishable feature when contextualizing a device.

Keifer et al explored and applied the findings made by Wanderley and Orio in a case study experiment on the usability of the Wii controller [4]. They found that whilst valuable data regarding their tested device's use as a music controller was insightful, they felt that their data was incomplete, as they did not measure the user's instantaneous musical experiences. Additionally, the concept of the 'third paradigm' in HCI was discussed in terms of DMI evaluation techniques. This paradigm is used to highlight the requirement for an ever-evolving selection of new evaluation techniques that suit the ubiquitous nature of computing in daily life. In essence, the third paradigm places embodied interaction at its centre. This means that all user actions, interactions, and knowledge are experienced and embodied within them and that they find meaning and construct meanings in specific contexts and situations [7].

Finally, a framework for evaluating DMIs was proposed by O'Modhrain in 2011 [8]. O'Modhrain examined the role of the various participants in the evaluation of the design process in a DMI context. At each stage in the design and development of the DMI the requisite participant (for example the inventor, manufacturer, musician) was given a formative role in the evaluation of a product's design. As such, we see the evaluation of a design taken from the perspective of an audience, performer/composer, designer, and manufacturer. The goal of each stakeholder is different and their means of assessment varies accordingly. That said, each perspective is necessary and should occur at different stages in an instrument's design cycle. O'Modhrain provides a conceptual scaffolding to bring together the various interested parties invested in the design process and explores the possibility of related or similar goals in an evaluation process.

## 4 Potential Assessment Techniques and Considerations

To accurately assess and compare DMIs, we must first categorize them to ensure their suitability for a particular test task. A general categorization can be made, identifying the basic elements of the instrument and the mechanical principles behind its operation. Following this, the characteristics of the DMI being analysed can be extended to include the physical variables involved in its manipulation. Developments in the taxonomy of input devices can be used to refine the classification variables down to two basic forms (force and position) and the derivatives found from the six possible degrees of freedom of each (translation and rotation in three directions) [9, 10]. These include the range of continuous and discrete values as generated by the DMI.

For the second step of our evaluation, we must contextualize our evaluation goals. This shifts the focus to the perspective of the evaluation process: who is evaluating and why? For this, we shall refer to the framework presented by O'Modhrain in 2011 [8]. Given the idiosyncratic design process carried out by most DMI designers, we suggest that evaluation goals from the viewpoint of performer/composer and designer be amalgamated in our evaluation framework. This is not to dismiss the perspectives of the audience or the manufacturer, but to highlight the role DMI designers often play as both the performer/composer and designer. Table 1 highlights aspects of device evaluation that best fits for these two stakeholder groups. From here on we should be able to draw upon existing HCI evaluation techniques and augment them to suit the chosen device's categorization, design taxonomy, and consideration of stakeholder requirements.

**Table 1.** Key Characteristics of Different Stakeholders in DMI Design Evaluation, extracted from O'Modhrain (2011) [8].

| | Possible Evaluation Goals | | | |
|---|---|---|---|---|
| Stakeholder | Enjoyment | Playability | Robustness | Achievement of Design Specifications |
| Performer / Composer | Reflective practice, development of repertoire, long-term engagement (longitudinal study) | Quantitative methods for evaluation of user interface, mapping, etc. | Quantitative methods for hardware / software testing | |
| Designer | Observation, questionnaire, informal feedback | Quantitative methods for user interface evaluation | | Use cases, feedback regarding stakeholder satisfaction |

After we have fully categorized, contextualized, and identified the stakeholders, we can consider HCI paradigms that are relevant to computing for specific applications. Given the current state of DMI evaluation, we cannot directly apply the same evaluation techniques as would be applicable to a Windows, Icons, Menus, and Pointers (WIMP) system, for instance. Nevertheless, we may still borrow from HCI techniques to assess a musical device's functionality, usability, and the musician's overall user experience.

In our evaluation, functionality shall refer solely to the technical capabilities of the device, making it possible to quantify what exactly the device does and how well it does it. This generally incorporates an analysis of the device's usefulness and reliability. In HCI, the characteristics of a usability analysis seek to measure the interaction between the user and the device in such a way as to ascertain if the device is capable of undertaking the tasks it is supposed to. It is important that prototype devices be as close to the final form factor, both in terms of design and functionality. Having a tangible working model of a device is key for a successful evaluation. Prototypes need to be functional, where gestures can be captured with precision, and in turn, they need to be responsive in sound generation without any noticeable latency. In a much broader sense, the measurement of a user's experience focuses on the relationship a user has with the device. This rests with the user's deeper emotional state in relation to a device, for example how they felt about their experience and if it meets their expectations of it as a musical device.

These three factors, although unique, do not operate independently of each other. For example, we do not consider usability as a defining device characteristic. However, the physicality of a device, in terms of its functionality and how it is delivered to the user, directly influences its usability. Also, a system's aesthetic beauty can influence the user's perception of usability and their physical experience with the device before actually using it. Finally, a device's usability directly influences the user's experience, as poor usability will almost certainly lead to a negative user experience. Therefore, we see the assessment of each of these areas is best achieved through the application of multiple HCI techniques and is not focused on any one alone.

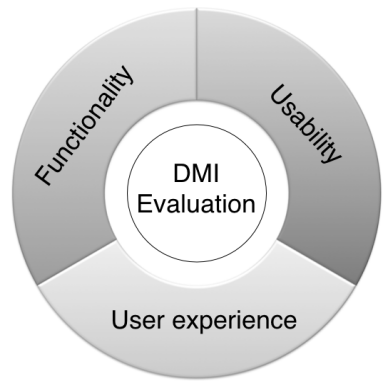

**Fig. 1.** A framework of DMI evaluation (adapted from [1]).

Functionality assessment is used to determine if the device's features afforded to the user are practical, as well as evaluating the performance, consistency, and the sturdiness of the designs used. To validate the functionality of a DMI, it must be capable of completing certain performance tasks, in other words, how it functions as a musical instrument. Additionally, we must also consider how a musician evaluates a device during performance. This includes their own subjective opinion of their performance, and the artistic freedoms afforded to them by the device must be measured. Therefore, a device that is being used to complete musical tasks for functionality testing must also include the incorporation of elements of usability and/or user experience in its analysis.

The musical tasks used to examine a device's effectiveness as an instrument, should be simple even if these tasks appear non-musical [11, 12]. This is due to the simple tasks being only the formative phase of a more complete device evaluation and should therefore not be considered in their entirety. Therefore, evaluation techniques such as Fitts' Law, Meyer's Law, and Steering Law [6], although basic and somewhat non-music centred in design, can be used to accurately measure and compare the functional aspects of a DMI.

Given the multiplicity of current DMI designs, we must carefully consider what it is that we are choosing to measure in order to evaluate the functionality of the designs. This is especially relevant to device comparison studies where the task must be achievable with all interfaces. A list of suggested musical tasks was made by Orio et al [6], as can be seen illustrated in Table 2. Additionally, we suggest some HCI evaluation techniques to test the device's functionality in these tasks. The outline presented in Table 2 is not representative of all musical tasks, and other HCI assessment techniques should also be considered. The breadth of both fields cannot be easily reduced to fit into so few categorical interactions, but the flexibility afforded in both can be manipulated to fit multiple conditions.

**Table 2.** Musical tasks can be linked with evaluation techniques from HCI.

| Musical Tasks | | Existing HCI Functionality Evaluation Methodologies |
|---|---|---|
| • Selecting an isolated tone: simple triggering to varying parameters such as pitch, loudness, and timbre.<br>• Musical gestures: glissandi, trills, grace notes, etc.<br>• Selecting scales and arpeggios at different speed, range, and articulation.<br>• Creating phrase contours: from monotonic to random.<br>• Ability to modulate timbre, amplitude or pitch for a given note and inside a phrase.<br>• Playing rhythms at different speeds and combining tones or pre-recorded materials.<br>• Synchronisation of musical processes. | 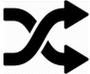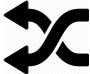Select an existing HCI methodology that best fits the musical task you wish to evaluate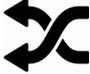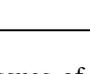 | • Target Acquisition - Fitts' Law.<br>• Pursuit Tracking - Control:Display ratio.<br>• Constrained Linear Motion Tracking.<br>• Constrained Circular Motion Tracking.<br>• Free-Hand Inking – subjective evaluation of facsimile of signature.<br>• Aimed movements composed as sub-movements - Meyer's Law.<br>• Measuring trajectory movements - Steering Law.<br>• Circular motion path tracking<br>• Varying trajectories path tracking |

Usability assessment is used to raise issues of efficiency, effectiveness, and user satisfaction. Further descriptions of device transparency, learnability, and feedback mechanisms can be drawn from analysing this data. The measure of usability is defined in ISO 9241-11 as '*quality in use*' [13]. Therefore, we should endeavour to reproduce this usability definition:

> "*… the extent to which a product can be used by specified users to achieve specific goals with effectiveness, efficiency, and satisfaction.*"

Beyond the ISO standard, there are a number of case studies that outline evaluation methodologies to assess a design's usability. However, care must be taken to choose an appropriate usability evaluation technique, which when applied to DMI devices, supports a high level of confidence in the findings. The chosen UEM must be capable of extrapolating the relevant information from the analysis. Known areas of concern include Learnability, Explorability, Feature Controllability, and Timing

Controllability [3]. We can expand upon these further and branch out the usability aspects of each to include other factors. These include:

a)  The demands a device places upon a user, such as cognitive load, physical exertion, temporal demands leading to fatigue, and

b)  How a device is *perceived* to affect a user's performance, the work involved in completing the task, and measuring frustration levels.

Learnability, as described in ISO 9241-11, is defined here as the time required to learn how to use the instrument. Learnability also incorporates the user's familiarity with the device or related devices, which is a difficult parameter to measure. However, a longitudinal study of usability should highlight learnability and playability issues that may arise from this. Findings should reflect the performer's previous training with specific instruments and their familiarity with other instruments within an ensemble. This information can be used to evaluate the amount of effort required to accomplish a task. Additionally, high levels of insecurity, discouragement, irritation, stress, and annoyance will reduce how much effort a performer will put into learning and applying the intricacies and nuances a device my bestow upon them. However, if a device is too easy to learn how to master, they will be as equally dissuaded from its use.

Explorability: the number of functions and capabilities afforded to the user and how they are implemented. We should be aware that all input parameters may be individually assessed for functionality or that assumptions can be made for inputs that share the same mechanical principles of operation. This should assist in the analysis of any multi-parametric control that is given, which is also dependent upon the mechanical characteristics of the chosen instrument.

Feature Controllability: the perceived accuracy, resolution, and range of the device. The ergonomic implications of a device's operation in terms of accuracy of movement, given the resolution and range of input gestures that are possible, allows designers and musicians/performers to evaluate if they have have fully achieved the capabilities of their design specifications. If they have not, users will evaluate their success in accomplishing a task negatively.

Timing Controllability: the fundamental difference between classical HCI observations and musical interactions is the central role of time. The measurement of input during a time-based exercise and its effect upon performance will give consideration to the simultaneous timing and rhythmic patterns that are central to musical performance. The temporal demands of a task should be achievable and flexible to a user's needs.

From this list of DMI considerations we can make use of a System Usability Scale (SUS) derived from existing HCI literature [14]. The SUS quickly outputs a number that represents a near instant measure of usability. Previously, we have seen such investigations successfully applied to psychometrically valid questioning for the evaluation of many products over the last 20 years. However, it can be argued that the standardized questions of a SUS analysis do not lend themselves to device evaluation in the 21ˢᵗ century. Therefore, we suggest that it may be necessary to augment and adapt them for the unique requirements of musical tasks. Further to this, we may also consider the NASA Task Load Index (NASA-TLX) as an effective measure of usability issues that are unique to DMIs [15]. This assessment technique has also been successfully applied many times to numerous studies that have provided a worthy resource for many usability-focused activities. Relating specifically to Learnability,

Explorability, Feature Controllability, and Timing Controllability, the NASA-TLX measures on a number of comparative scales. The scales of the NASA-TLX measure the following demands; Mental, Physical, Temporal, Performance, Effort, and Frustration Level. Using this set of six subscales, the overall workload can be analysed in order to extrapolate information pertaining to the individual factors of Learnability, Explorability, Feature Controllability, and Timing Controllability. The definition of each subscale can be seen in Table 3.

**Table 3.** NASA-TLX rating scale definitions extracted from Hart (1988) [21].

| Subscale | Description |
| --- | --- |
| Mental: | How much mental and perceptual activity was required? Was the task easy or demanding, simple or complex, exacting or forgiving? |
| Physical: | How much physical activity was required? Was the task easy or demanding, slow or brisk, slack or strenuous, restful or laborious? |
| Temporal: | How much time pressure did you feel due to the rate or pace at which the task elements occurred? Was the pace slow and leisurely or rapid and frantic? |
| Performance: | How successful do you think you were in accomplishing the goals of the task set by the experimenter? How satisfied were you with your performance in accomplishing these goals? |
| Effort: | How hard did you have to work to accomplish your level of performance? |
| Frustration: | How insecure, discouraged, irritated, stressed and annoyed versus secure, gratified, content, relaxed and complacent did you feel during the task? |

Each aspect of usability in HCI can be analysed independently. Specifically, efficiency, effectiveness, and user satisfaction data can be collected from a combination of different sources. Efficiency can be established by measuring the mental effort required to use the DMI: for example, a low mental effort would indicate a high efficiency in operation. This data can be collected using a modified post-task self-report Subject Mental Effort Questionnaire (SMEQ) [16] and a Single Ease Question (SEQ) [17]. We can also use data collected from functionality testing to ascertain device effectiveness. Functionality data can be supported with additional usability studies. Finally, the satisfaction of the participant can be measured using a modified Consumer Product Questionnaire (CPQ) [2]. In order for a researcher to address the areas of concern outlined earlier, they can modify each of these styles of HCI usability testing. Additionally, they may also attain specific knowledge depending upon the device being tested and the overall aims of the research being undertaken.

Assessing a user's experience is a relatively new and innovative area of investigation within the field of HCI. A number of appraisal methodologies exist, but they remain under-developed due to still being in the early stages of creation. Additionally, the evocative nature of the relationship a musician develops with certain types of musical instruments can be idiosyncratic and diverse in its formative stages. Moreover, any data collected in user experience testing is altogether subjective in nature. Measurements are difficult to quantify and can be dependent on a number of contributing influences, such as psychological or social factors [18]. An example might include personal opinions on aesthetics, a user's exposure to advertising, or the social desirability of certain technologies. User experience can be measured in a number of ways, but we will focus on the adaption of three particular methods. Firstly, we suggest that a simple preference report can be used to summarize comparisons with other devices. Secondly, a User Experience Questionnaire (UEQ) can be conducted [19]. Finally, we propose that qualitative data should be collected relating to the contributing factors of a musician's experience whilst performing both functionality and musical tasks by using a Critical Incidents Technique (CIT) analysis [20]. The adaptation of these techniques serves to provide a flexible, yet validated and constrained, user experience measure for comparison.

## 5 Conclusion

Models of evaluation exist in both fields of DMI design and HCI that can serve as guidelines for future DMI appraisals. Currently, DMIs are often evaluated idiosyncratically, and established evaluation methods from other areas are somewhat ignored. We have investigated and presented several existing methods of device evaluation from both areas that may be applied to new musical interfaces. Specifically, we have highlighted a number of steps to ensure that a complete and in-depth device appraisal is carried out. In our device appraisal conclusions, we stress the need for established, rigorous, and flexible techniques. The field of HCI contains many validated techniques that have been successfully applied over many years. However, the evaluation of a musical device is often far more complex in practice than a conventional computer interface or device. Therefore, experimentation must be undertaken to find an appropriate evaluation technique that best fits the device.

The initial stages of a device's evaluation should include the capture of low-level device characteristics, creating a generalized device description. Following this, a device should be reduced to its physical variables in terms of taxonomy of input. The second step here should contextualize a device's evaluation in terms of stakeholder, questioning who is evaluating the device and why. These initial steps serve to inform the evaluation and comparison of functionality studies that follow. Devices are required to be capable of undertaking the analysis tasks and must be analogous in operation if they are to be compared. A variety of potential HCI paradigms exist that can be augmented to best fit the categorization and contextualization outlined in the first stage. The main categories to measure include a device's functionality, usability, and the user's experience with the device. Functionality testing should also include an element of analysis of the usability and user experience, as functionality testing is able to highlight any potential issues in this area before longitudinal studies are carried out. Usability and user experience in a musical context requires a longitudinal study; as musicians must be given time to evaluate a device in a natural setting over time. The application of multiple HCI questioning techniques will highlight important usability and user experience data in a real-world application of the device.